\documentclass[prc,twocolumn,showpacs,amsmath,amssymb]{revtex4}
\usepackage{times}
\usepackage{graphicx}
\usepackage{dcolumn}
\usepackage{bm}
\usepackage{amstext}

\begin{document}

\title{The evaporation residue in the fission state of Barium nuclei within relativistic mean-field theory}
\author{M. Bhuyan$^1$, S. K. Patra$^1$ and Raj K. Gupta$^2$}
\affiliation{$^1$ Institute of Physics, Sachivalaya Marg, Bhubaneswar- 751 005, India. \\ 
$^2$ Department of Physics, Panjab University, Chandigarh- 160 014, India.}

\email{bunuphy@iopb.res.in}

\date{\today}

\begin{abstract}

The evaporation residue of Barium isotopes are investigated in a microscopic study using relativistic mean field theory. 
The investigation includes the isotopes of Barium from the valley of stability to exotic proton-rich region. The ground 
as well as neck configurations for these nuclei are generated from their total nucleonic density distributions of the 
corresponding state. We have estimated the constituents (number of nucleons) in the elongated neck region of the fission 
state. We found the $\alpha$-particle as the constituent of neck of Ba-isotopes, referred to as the evaporated residue 
in heavy-ion reaction studies. A strong correlation between the neutron and proton is observed throughout the isotopic 
chain.

\end{abstract}


\maketitle

\section{Introduction}
In atomic nuclei, the occurrence of clusters and their cluster structures were already predicted in late 1930's by Wheeler
and von Weizacker \cite{wheel37,wiss38}. Numerous experimental and theoretical studies based on advanced microscopic 
approaches with single-nucleon degrees of freedom have revealed a wealth of data on clustering phenomena in light nuclei 
\cite{free06,kanada01,biel95,neff03,arum05,gupta08a}. Another instance of clustering in atomic nuclei is the cluster 
radioactivity, first predicted theoretically \cite{sandu80,peon91}, based on quantum mechanical fragmentation theory 
\cite{fink74,gupta77,gupta89}, and then observed experimentally \cite{rose84}. Few other testimony for the formation of 
nuclear clustering are reported from experiments, but indirectly. The most possible clusters in the ground and excited 
states of light and medium mass nuclei are the $N=Z$ $^4$He, $^8$Be, $^{12}$C, $^{16}$O, $^{20}$Ne, $^{24}$Mg and $^{28}$Si 
nuclei \cite{kanada01,suzu04,waka07,le10,kumar94,kumar95}.

Clustering is the essential feature of many-nucleon dynamics that coexists with the nuclear mean-field. In this context,
one can say that a fully microscopic understanding and description of cluster formation and their emission is necessary in 
mean-field approaches, despite the existence of clusterization effects in many calculations \cite{bhu11a,patra11,bhu13}. 
This is because, in most of these calculations, we are neglecting the structure effects such as the binding energies and 
scattering phase shifts of these configurations that are assumed as a-priory ingredients of effective interactions in the 
mean-field models \cite{bhu09,bhu11,bhu13a}. Another consequence is that the nuclear deformation in ground state plays an 
important role in the formation of a cluster because it removes the degeneracy of single nucleon levels associated with 
spherical symmetry. At some specific deformation, the restored degeneracy degenerates the deformed shell closures and 
facilitates the formation of clusters. However, this is a rather qualitative explanation because the clustering phenomenon 
can not generally be explained by accidental degeneracies. Hence, in order to understood the mechanism of clustering, a 
more general description that encompasses both cluster and quantum liquid-drop aspects in the finite nuclear systems are 
needed.

The aim of this work is to address the origin of neck, assumed to be required for the preformed clusters in the fission 
state, i.e., the formation of elongated neck in the fission state of a nucleus. It is worth mentioning that the 
relativistic mean-field approach is able to give a complete and accurate description of ground-state properties and 
collective excitations over the whole nuclear chart including the superheavy mass regions \cite{bhu11a,bhu09,bhu11}. In 
this approach, the many-body dynamics is represented by independent nucleons interacting through the meson field with a 
local self-consistent mean-field potential that corresponds to the actual density and current distribution of a given 
nucleus for a given state. In a recent work \cite{bhu11a}, we have explained the formation of cluster(s) and their 
composition in the ground and intrinsic excited state(s) of Ba isotopes. In the present study, we are interested to 
reveal the fission state of Ba nuclei from a microscopic description based on the framework of relativistic mean field 
theory. The most important steps that have been taken here are to find the constituents or the composition of neck, which 
plays an important role for the explanation of a fission state. In a fission process, the neck is well accepted to be the 
nucleon's emission region \cite{patra11,bhu11a}. It will truly be of benefit if it would be possible to generate the neck 
structure theoretically and find out the composition of this region quantitatively. Such a study will be accessible for 
understanding of the fission state as well as its residue in an heavy-ion collision experiment where the excited compound 
nucleus decays by the emission of light-particle evaporation residues and the fission fragments. For Ba$^*$ compound 
systems (specifically, $^{118,122}$Ba$^*$), though the charge distribution for fission fragments with 3$\le$Z$\le$28 is 
measured \cite{bonnet08,ademard11}, only the total cross-section $\sigma_{ER}$ is measured for the evaporation residues 
(ER: Z$<$3)), but without identifying the constituents (Z or A-value) of $\sigma_{ER}$. It is, therefore, of interest to 
know the constituent(s) or the particle(s) in the neck region of the fission state of Ba nuclei, in the RMF formalism - a 
structure calculation. These constituent(s) will belong to ER or fission, depending on the charge number (Z-value) of the 
particle(s) in neck region is less than or greater than Z=2.

The paper is organized as follows: Section II gives a brief description of the relativistic mean-field formalism (RMF). The 
effects of pairing for open shell nuclei, included in our calculations, are also discussed in this section. The results of 
our calculations are presented in Section III. The neck configuration and their composition are also discussed 
quantitatively in this section. A summary of the results obtained, together with concluding remarks, are given in the last 
Section IV.

\section{The relativistic mean-field Theory}
In the last few decades, the relativistic mean field theory is applied successfully to study the structural properties of 
nuclei throughout the nuclear chart, including also the unknown island of superheavy nuclei \cite{bhu09,sero86,boguta77,
horo81,boguta83,patra91,ring86}. The relativistic Lagrangian density for nucleon-meson many-body system is expressed as 
\cite{boguta83,ring86},
\begin{eqnarray}
{\cal L}&=&\overline{\psi_{i}}\{i\gamma^{\mu}
\partial_{\mu}-M\}\psi_{i}
+{\frac12}\partial^{\mu}\sigma\partial_{\mu}\sigma
-{\frac12}m_{\sigma}^{2}\sigma^{2}\nonumber\\
&& -{\frac13}g_{2}\sigma^{3} -{\frac14}g_{3}\sigma^{4}
-g_{s}\overline{\psi_{i}}\psi_{i}\sigma-{\frac14}\Omega^{\mu\nu}
\Omega_{\mu\nu}\nonumber\\
&&+{\frac12}m_{w}^{2}V^{\mu}V_{\mu}
+{\frac14}c_{3}(V_{\mu}V^{\mu})^{2} -g_{w}\overline\psi_{i}
\gamma^{\mu}\psi_{i}
V_{\mu}\nonumber\\
&&-{\frac14}\vec{B}^{\mu\nu}.\vec{B}_{\mu\nu}+{\frac12}m_{\rho}^{2}{\vec
R^{\mu}} .{\vec{R}_{\mu}}
-g_{\rho}\overline\psi_{i}\gamma^{\mu}\vec{\tau}\psi_{i}.\vec
{R^{\mu}}\nonumber\\
&&-{\frac14}F^{\mu\nu}F_{\mu\nu}-e\overline\psi_{i}
\gamma^{\mu}\frac{\left(1-\tau_{3i}\right)}{2}\psi_{i}A_{\mu} .
\end{eqnarray}
All the quantities in the above Lagrangian density have their usual well known meanings. From the Lagrangian we obtain the 
field equations for the nucleons and mesons. These equations are solved by expanding the upper and lower components of the 
Dirac spinors and the boson fields in an axially deformed harmonic oscillator basis with an initial deformation 
$\beta_{0}$. The set of coupled equations is solved numerically by a self-consistent iteration method. The center-of-mass 
motion energy correction is estimated by the usual harmonic oscillator formula $E_{c.m.}=\frac{3}{4}(41A^{-1/3})$. The 
quadrupole deformation parameter $\beta_2$ is evaluated from the resulting proton and neutron quadrupole moments, as 
$Q=Q_n+Q_p=\sqrt{\frac{16\pi}5} (\frac3{4\pi} AR^2\beta_2)$. The root mean square (rms) matter radius is defined as 
$\langle r_m^2\rangle={1\over{A}}\int\rho(r_{\perp},z) r^2d\tau$, where $A$ is the mass number, and $\rho(r_{\perp},z)$ is 
the deformed density. The total binding energy and other observables are also obtained by using the standard relations, 
given in Ref. \cite{ring90}. We use the well known NL3 parameter set \cite{lala97}, which not only reproduces the 
properties of stable nuclei but also well predicts for those nuclei far from the $\beta$-stability valley. As outputs, we 
obtain different potentials, densities, single-particle energy levels, radii, deformations and the binding energies. 
For a given nucleus, the maximum binding energy corresponds to the ground state and other solutions are obtained as various 
excited intrinsic states, provided the nucleus does not undergo fission. For the fission state \cite{rutz95,rutz98,book01}, 
we follow a special procedure discussed below. 

\subsection{Pairing Effect}

In the present study, we are dealing with the proton-rich isotopes of Ba nucleus, hence pairing is a crucial quantity for 
the open shell nuclei in determining the nuclear gross properties. The constant gap, BCS-pairing approach has been adopted 
for the present calculations. The general expression for pairing energy is given as:
\begin{eqnarray}
E_{pair}=-G\left[\sum_{i>0}u_{i}v_{i}\right]^2, 
\end{eqnarray} 
where $G$ is the pairing force constant and $v_i^2$ and $u_i^2=1-v_i^2$ are the occupation probabilities. The variational 
procedure with respect to the occupation numbers $v_i^2$, gives the BCS equation 
$2\epsilon_iu_iv_i-\triangle(u_i^2-v_i^2)=0$ with $\triangle=G\sum_{i>0}u_{i}v_{i}$. This is the famous BCS equation for 
pairing energy. The densities are contained within the occupation number 
$n_i=v_i^2=\frac{1}{2}\left[1-\frac{\epsilon_i-\lambda}{\sqrt{(\epsilon_i-\lambda)^2+\triangle^2}}\right].$
In order to take care of the pairing effects in the present study, we use the constant gap for proton and neutron, as given 
in \cite{madland81}:  
$\triangle_p =RB_s e^{sI-tI^2}/Z^{1/3}$ and $\triangle_n =RB_s e^{-sI-tI^2}/A^{1/3}$, with $R$=5.72, $s$=0.118, 
$t$= 8.12, $B_s$=1, and $I = (N-Z)/(N+Z)$. (Note that the gaps obtained by these expressions are valid for nuclei both on 
or away from $\beta$-stable region.) 

In solving the RMF equations, the pairing force constant $G$ is not calculated explicitly. Instead, using the above gap 
parameter, we calculate directly the occupation probability using the chemical potentials for nucleons $\lambda_n$ and 
$\lambda_p$. Finally, we can write the pairing energy as:
\begin{eqnarray}
E_{pair}=-\triangle\sum_{i>0}u_{i}v_{i}^2. 
\end{eqnarray}
Apparently, in a given nucleus, for a constant pairing gap $\triangle$, the pairing energy $E_{pair}$ is not constant since 
it depends on the occupation probabilities $v_i^2$ and $u_i^2$, and the deformation parameter $\beta_2$. For example, for a 
constant pairing parameter $\triangle$ and force constant $G$, the pairing energy $E_{pair}$ diverges if it is extended to 
an infinite configuration space. In fact, in all realistic calculations with finite range forces, $\triangle$ decreases 
with state (spherical or deformed) for large momenta near the Fermi surface. However, in the present case, we assume that 
pairing gap for all states $\mid\alpha>=\mid nljm>$ are equal to each other near the Fermi surface and hence a constant
pairing gap is taken for simplicity of the calculations. We use in our calculations a pairing window, and all the equations 
extended up to the level $\epsilon_i-\lambda\leq 2(41A^{1/3})$. The factor 2 has been determined so as to reproduce the 
pairing correlation energy for neutrons in $^{118}$Sn using Gogny force \cite{ring90}. This type of prescription for 
pairing effects, both in RMF and Skyrme Hartree Fock (SHF), has already been used by us and many others authors 
\cite{patra01,bhu11a}. Within this pairing approach, it is shown \cite{patra01,lala99} that the results for binding 
energies and quadruple deformations are almost identical with the predictions of relativistic Hartree-Bogoliubov (RHB) 
approach \cite{lala99,lala01}.

\begin{table*}
\caption{The RMF(NL3) results for binding energy, pairing energy, the charge radii and quadrupole deformation parameter 
$\beta_2$ for $^{112-134}$Ba nuclei for the ground-state and fission state, compared with the FRDM predictions 
\cite{moll95,moll97} and the experimental data \cite{audi13,angeli13,raman01}. The energy in $MeV$ and radius in $fm$.}
\renewcommand{\tabcolsep}{0.15cm}
\renewcommand{\arraystretch}{1.3}
\begin{tabular}{ccccccccccc}
\hline \hline
Nucleus&\multicolumn{3}{c}{Binding Energy}& Pairing Energy &\multicolumn{3}{c}{Charge Radius}
& \multicolumn{3}{c}{Deformation Parameter} \\
\hline
& NL3 & Expt. & FRDM & $E_{pair}$& NL3 & Expt. & FRDM & NL3 & Expt. & FRDM \\
\hline
$^{112}$Ba &895.33 &       &894.88 &21.43&4.754&      &&0.239&     &0.207\\
   &857.12 &       &       &19.35&9.251&      &&5.562&     &     \\
$^{114}$Ba&920.67 &922.26 &921.26 &21.29&4.765&      &&0.235&     &0.243\\
   &880.37 &       &       &19.85&9.252&      &&5.561&     &     \\
$^{116}$Ba&947.55 &947.14 &946.85 &20.82&4.786&      &&0.294&     &0.280\\
   &902.14 &       &       &20.15&9.255&      &&5.564&     &     \\
$^{118}$Ba&971.43 &970.90 &970.74 &19.89&4.805&      &&0.330&     &0.290\\
   &922.51 &       &       &20.13&9.266&      &&5.564&     &     \\
$^{120}$Ba&994.09 &993.63 &993.43 &19.55&4.810&4.8092&&0.320&     &0.281\\
   &935.31 &       &       &19.51&9.494&      &&5.574&     &     \\
$^{122}$Ba&1015.69&1015.50&1015.20&18.92&4.816&4.8153&&0.321&0.345&0.273 \\
   &953.92 &       &       &18.65&9.505&      &&5.575&     &      \\
$^{124}$Ba&1036.17&1036.12&1035.98&18.31&4.822&4.8185&&0.295&0.302&0.274 \\
   &978.25 &       &       &18.20&9.300&      &&5.577&     &      \\
$^{126}$Ba&1055.66&1055.84&1055.67&17.66&4.820&4.8221&&0.252&0.273&0.256 \\
   &992.86 &       &       &18.46&9.293&      &&5.572&     &      \\
$^{128}$Ba&1074.77&1074.68&1074.22&16.82&4.820&4.8255&&0.215&0.249&0.218 \\
   &1005.77&       &       &18.71&9.306&      &&5.578&     &      \\
$^{130}$Ba&1093.19&1092.72&1092.04&16.02&4.823&4.8283&&0.181&0.218&0.171 \\
   &1019.91&       &       &17.27&9.338&      &&5.585&     &      \\
$^{132}$Ba&1110.89&1110.04&1109.24&15.24&4.826&4.8303&&0.143&0.186&0.143 \\
   &1034.57&       &       &15.69&9.351&      &&5.581&     &      \\
$^{134}$Ba&1127.69&1126.70&1126.13&14.38&4.830&4.8322&&0.101&0.161&-0.113\\
   &1049.79&       &       &14.39&10.48&      &&5.583&     &      \\
\hline \hline
\end{tabular}
\label{Table 1}
\end{table*}

\section{Details of the calculations and Results}

To-date, there exists a large number of force parameters for finding a convergent solution of the RMF Lagrangian density. In 
many of our previous works and of others \cite{patra91,ring90,lala97,bhu09,bhu11a,bhu11,bhu13} the ground state properties 
such as the binding energies (BE), quadrupole deformation parameters $\beta_2$, root-mean-square charge radii ($r_{ch}$), 
single-particle level, nucleonic distributions and other bulk properties, are evaluated by using the various parameter 
sets. From these works, one can conclude that, more or less, most of the recent parameter sets reproduce well the ground 
state properties, not only of stable normal nuclei but also for exotic nuclei including superheavy. This implies that by 
using one reasonably acceptable parameter set, the predictions of the model will remain nearly force independent. Here, we 
have used the most popular NL3 force parameter for this investigation. The numerical calculations are carried out by taking 
the maximum oscillator quanta $N_F=N_B=16$ for Fermion and boson. To test the convergence of the solutions, few 
calculations are done with $N_F$=$N_B$=12 also. The variation of these two solutions are $\leq 0.002\%$ for binding energy 
and 0.001$\%$ for nuclear radii in drip-line nuclei. Hence, the used model space is good enough for the considered nuclei 
in a large deformed state. The number of mesh points for Gauss-Hermite and Gauss-Lagurre integration are 20 and 24, 
respectively. For a given nucleus, as already stated above, the maximum binding energy corresponds to the ground-state, and 
the other solution with very high deformation is for the fission state. 

\subsection{Binding energy, nuclear charge radius and deformation parameter}

To deal with the fission state of a nucleus, it is important as well as necessary to know their ground-state bulk 
properties, which are mostly responsible for the internal configuration of nuclei. In this context, we first calculated the 
gross properties such as binding energy, rms charge radius $r_{ch}$ and quadrupole deformation parameter $\beta_2$ of 
Ba-isotopes by using the RMF with NL3 parameter set. The obtained results from RMF(NL3) are compared with the Finite Range 
Droplet Model (FRDM) predictions \cite{moll95,moll97} and the experimental data \cite{audi13,angeli13,raman01}, wherever 
possible. All the results are listed in Table I. From the table, we find that, in general, the microscopic binding 
energies, rms charge radii $r_{ch}$ and deformation parameters agree well with the experimental data. The closer comparison 
of RMF with the FRDM and experimental data shows clearly that the binding energy and radius coincide remarkably well 
throughout the isotopic chain. However, in case of deformation parameter $\beta_2$, the RMF matches the experimental data 
but not the FRDM predictions. For example, the RMF shows prolate deformed shapes for all $^{122-134}$Ba isotopes, which 
compare nicely with the experimental data but not with the FRDM results, in particular for $^{134}$Ba where it is predicted 
to be oblate in FRDM. 

%%%%%%%%%%%%%%%%%%%%%%%%%%%%%%%%%%%%%%%
\begin{figure}
%\vspace{0.6cm}
\begin{center}
\includegraphics[width=1.25\columnwidth]{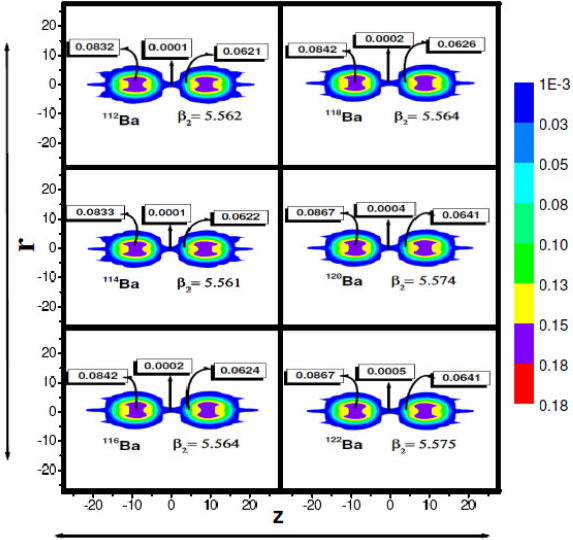}
\caption{The neck structure for $^{112-122}$Ba in axially deformed coordinates for RMF calculations using NL3 parameter 
set.
}
\end{center}
\label{Fig. 1}
\end{figure}
%%%%
\begin{figure}[ht]
%\vspace{0.55cm}
\begin{center}
\includegraphics[width=1.25\columnwidth]{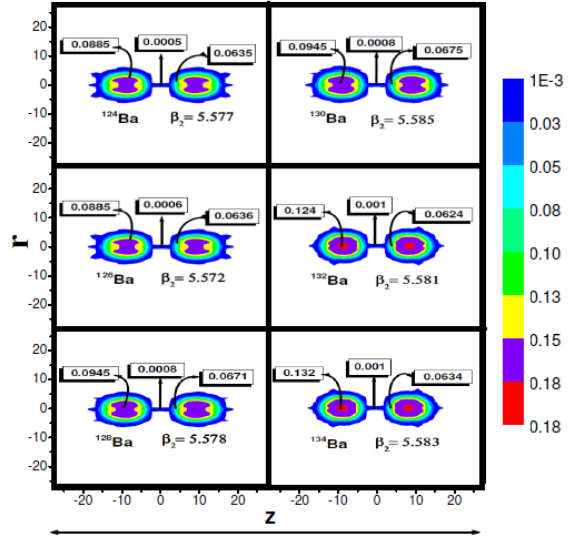}
\caption{Same as for Fig. 1, but for $^{124-134}$Ba.
}
\end{center}
\label{Fig. 2}
\end{figure}
%%%%%%%%%%%

\subsection{The neck configurations of $^{112-134}$Ba}

Generally, the internal configuration of a nucleus can be determined from the density distributions of the nucleons for a 
given state. The total density of a nucleus (i.e. sum of the neutron and proton densities) in the $\rho z$ plane from 
RMF(NL3) are obtained for the positive quadrant of the plane parallel to the symmetry $z-$axis. Here, $\rho=x=y=r_{\bot}$ 
and $z$ is the symmetric axis. It is to be noted that, both the axes $z$ and $\rho$ are conserved in the present formalism 
under the space reflection symmetry. Now we can obtain the complete picture of a nucleus in the $\rho-z$ plane by 
reflecting the first quadrant to other successive quadrants. The contour plotting of density along with the color code with 
corresponding density ranges for the neck structure of the fission state of $^{112-134}$Ba are shown in Figs. 1 and 2. 
From the color code, one can identify the magnitude of the density range of a particular color code. For example, the color 
code with deep red corresponds to maximum density $\rho \sim$ 0.15 $fm^{-3}$ and the olive bearing the minimum value of 
$\rho \sim$ 0.001 $fm^{-3}$. (In black and white figures, the color code is read as deep black with maximum density 
to outer gray as minimum density distribution). 
%%%%%%%%%%%%%%%%%%%%%

\begin{table}
\caption{The RMF(NL3) results for neck configurations of $^{112-134}$Ba in the fission state giving the ranges of the neck 
and their Z and N constituents.}
\renewcommand{\tabcolsep}{0.1cm}
\renewcommand{\arraystretch}{1.2}
\begin{tabular}{cccccccccccc}
\hline \hline
Nucleus& $\beta_2$ & Range ($r_1$,$r_2$; $z_1$,$z_2$) & $Z_{neck}$
& $N_{neck}$ & Neck Nucleus \\
\hline
$^{112}$Ba & 5.562& ($\pm 2.28$; $\pm 1.25$) & 2.07 & 1.91 & $^{4}$He \\
$^{114}$Ba & 5.561& ($\pm 2.28$; $\pm 1.25$) & 2.04 & 2.95 & $^{4}$He \\
$^{116}$Ba & 5.564& ($\pm 2.27$; $\pm 1.25$) & 2.05 & 2.03 & $^{4}$He \\
$^{118}$Ba & 5.564& ($\pm 2.27$; $\pm 1.26$) & 2.06 & 2.13 & $^{4}$He \\
$^{120}$Ba & 5.574& ($\pm 2.26$; $\pm 1.26$) & 2.08 & 2.08 & $^{4}$He \\
$^{122}$Ba & 5.575& ($\pm 2.26$; $\pm 1.26$) & 2.05 & 2.18 & $^{4}$He \\
$^{124}$Ba & 5.577& ($\pm 2.26$; $\pm 1.26$) & 2.05 & 2.08 & $^{4}$He \\
$^{126}$Ba & 5.572& ($\pm 2.26$; $\pm 1.27$) & 2.09 & 2.08 & $^{4}$He \\
$^{128}$Ba & 5.578& ($\pm 2.26$; $\pm 1.27$) & 2.11 & 1.98 & $^{4}$He \\
$^{130}$Ba & 5.585& ($\pm 2.25$; $\pm 1.27$) & 2.09 & 2.18 & $^{4}$He \\
$^{132}$Ba & 5.581& ($\pm 2.25$; $\pm 1.27$) & 2.09 & 2.13 & $^{4}$He \\
$^{134}$Ba & 5.583& ($\pm 2.25$; $\pm 1.27$) & 2.08 & 1.12 & $^{4}$He \\
\hline \hline
\end{tabular}
\label{Table 2}
\end{table}
%%%%%%%%%%%%%%%%%%%%%%%%

\begin{figure}
%\vspace{-0.6cm}
\begin{center}
\includegraphics[width=1.15\columnwidth]{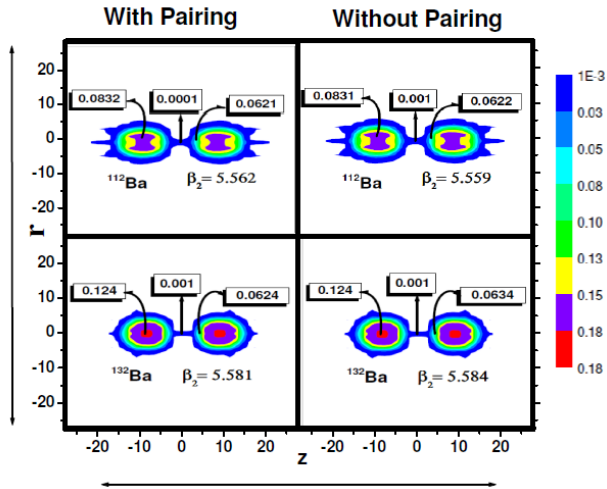}
\caption{Same as for Fig. 1, but for $^{112,134}$Ba, with and without pairing.
}
\end{center}
\label{Fig. 3}
\end{figure}
%%%%%%%%%%%%%%%%%%%%%%%%%%%%%%%%%%%%%%

\subsection{The neck structure with and without pairing}

The pairing is important for open shell nuclei, near and far away from $\beta$-stable region of the nuclear chart. However, 
for a given nucleus, its value depends slightly on the binding energy and marginally on quadrupole deformation $\beta_2$. 
This means to say that for differing $\beta_2$-values in a nucleus, the pairing energy $E_{pair}$ changes only marginally 
($\sim$2-5$\%$ for this specific region). On the other hand, even if the $\beta_2$ values for two nuclei are same, 
$E_{pair}$ values are different from one another, depending on the filling of the nucleons. This result is illustrated
in Table I for the RMF(NL3) calculation, where pairing energy $E_{pair}$ for both the ground-state ($g.s.$) and fission 
state ($f.s.$), and their corresponding $\beta_2$-values, are displayed. A careful inspection of the pairing energy shows 
a linear decrease in the magnitude from drip-line to $\beta$-stable line. This trend is also valid for the fission states 
of these nuclei. For example, the magnitude of pairing energy $E_{pair}$ for $^{112}$Ba and $^{134}$Ba are 21 and 14 MeV, 
respectively. The relative difference in the two $E_{pair}$ values is 7 MeV, i.e., the change of $E_{pair}$ from 
drip-line to $\beta$-stability line is $\sim$30$\%$ for the isotopic chain of Ba. It is well known that the RMF formalism 
reproduce the cluster structure of nuclei \cite{bhu11a,bhu13,patra11} which are already predicted by several cluster models 
\cite{hori01,funa03,funa06}. Here, the nucleons are treated as point particles, which oscillates in the mean field of meson 
medium and gives the way for an independent constraint, resulting in clustering inside the nucleus. On the other hand, one 
can say that the fluctuations in the central density of the nucleons are due to the shell effects, which may be caused by 
the pairing correlation of nucleons around the Fermi surface. If this is true, then it would be hard to find the shell gaps 
and other magic properties in the RMF formalism. Furthermore, we have examined the pairing effect on the neck structure of 
nuclei. For this purpose, we have plotted the contour of density profiles of $^{112,134}$Ba in their fission state, with 
and without pairing, as the representative cases shown in Fig. 3. From the figure, we found almost identical structures in 
both the cases. Hence, we one can say that the clustering inside the nuclei is almost independent or very slightly 
dependent on pairing.

\subsection{The neck and its constituents}

A careful inspection of the Figs. 1-3 show the formation of a neck structure inside the nucleus, i.e., the preliminary 
stage of fission. This region has very high decay probability than the other interior part of the nucleus. The most 
important and crucial attempt made by us here is to see the constituents (neutrons and protons) in this region. 

Basically, the constituent of a region depends on the size and the magnitude of the density for that area. To determine the 
neck structure inside the nucleus, it is important to know the volume of that region, i.e., the ranges of the elongated 
area in the form of a neck. It is worth mentioning here that the ranges are fixed by graphical method which is guided 
by the eyes and may have $1-2\%$ uncertainty in the results. The ranges for the necks of $^{112-134}$Ba, calculated from 
Figs. 1-3, are given in Table II. The formula used to identify the ingredient of the neck in the fission state is given 
by \cite{bhu11a,bhu13,patra11}:
\begin{equation}
n=\int_{z_1}^{z_2}\int_{r_1}^{r_2}\rho (z,r_{\bot})d\tau,
\end{equation}
where, $n$ is the number of neutrons $N$ or protons $Z$ or mass $A$, and $z$ ($z_1$, $z_2$) and $r_{\bot}$ ($r_1$ , $r_2$) 
are the ranges. From the estimated proton and neutron numbers, we determine the mass number of the nucleus emitted at the 
time of fission. The obtained nucleons for the fission state, along with the corresponding deformation, for Ba-isotopes are 
listed in Table II. From the table, we notice the presence of an $\alpha$ (i.e., $^{4}$He) as the residue in Ba-isotopes. 
Though a straight forward calculation, for the fission state, this is being carried out here for the first time.

\section{Summary and Conclusions}

Concluding, we have presented the gross nuclear properties, like the binding energy, deformation parameter $\beta_2$, 
charge radius $r_{ch}$ and the nucleonic density distributions for the isotopic chain $^{112-134}$Ba, using an axially 
deformed relativistic mean field formalism with NL3 parameter set. The results of our calculations show a quantitative 
agreement with the experimental data. We found prolate deformed ground-state solutions for Ba isotopes, which are 
consistent with the experimental data but do not match the FRDM predictions for some cases. 

Analyzing the total nuclear density distributions, the neck structure, i.e., the fission states of Ba-isotopes are 
identified. The effect of pairing on the neck structure are taken into account. We find that the neck of Ba-isotopes is
built up of $^{4}$He nuclei, which can be taken as the evaporation residue of their decay process. Experimentally, the 
evaporation residue of a decaying compound system consist of multiple neutrons, protons and $\alpha$-particle. It is 
further noticed that the neck structure of a nucleus remains unaffected for different force parameters, as long as the 
solutions for that nucleus exist. It will be interesting to measure the constituents of evaporation residues of hot 
Ba-isotopes formed in heavy-ion reactions.

\end{document}